\documentclass[aps,preprint,preprintnumbers,nofootinbib,superscriptaddress,tightenlines,byrevtex,showpacs]{revtex4}
\usepackage{amsmath,amssymb,amsbsy}
\usepackage{graphicx,subfigure}
\usepackage{comment}

\def\str{{\mathrm{str}}}

\begin{document}

\unitlength=1mm

\def\a{{\alpha}}
\def\b{{\beta}}
\def\d{{\delta}}
\def\D{{\Delta}}
\def\e{{\epsilon}}
\def\g{{\gamma}}
\def\G{{\Gamma}}
\def\k{{\kappa}}
\def\l{{\lambda}}
\def\L{{\Lambda}}
\def\m{{\mu}}
\def\n{{\nu}}
\def\w{{\omega}}
\def\O{{\Omega}}
\def\S{{\Sigma}}
\def\s{{\sigma}}
\def\t{{\tau}}
\def\th{{\theta}}
\def\x{{\xi}}

\def\ol#1{{\overline{#1}}}

\def\Dslash{D\hskip-0.65em /}
\def\dslash{{\partial\hskip-0.5em /}}
\def\vslash{{\rlap \slash v}}
\def\qbar{{\overline q}}

\def\CPT{{$\chi$PT$\;$}}
\def\QCPT{{Q$\chi$PT}}
\def\PQCPT{{PQ$\chi$PT}}
\def\tr{\text{tr}}
\def\str{\text{str}}
\def\diag{\text{diag}}
\def\order{{\mathcal O}}
\def\vit{{\it v}}
\def\vD{\vit\cdot D}
\def\am{\alpha_M}
\def\bm{\beta_M}
\def\gm{\gamma_M}
\def\smb{\sigma_M}
\def\smt{\overline{\sigma}_M}
\def\tb{{\tilde b}}
\def\mp{{m_\pi}}
\def\fp{{f_\pi}}
\def\fps{{f_\pi^s}}
\def\fpt{{f_\pi^t}}
\def\delmd{{\partial_\mu}}
\def\delmu{{\partial^\mu}}
\def\delnd{{\partial_\nu}}
\def\delnu{{\partial^\nu}}

\def\mc#1{{\mathcal #1}}

\def\Bbar{\overline{B}}
\def\Tbar{\overline{T}}
\def\cBbar{\overline{\cal B}}
\def\cTbar{\overline{\cal T}}
\def\pq{(PQ)}

\def\eqref#1{{(\ref{#1})}}

%
%
\newcount\hour \newcount\hourminute \newcount\minute 
\hour=\time \divide \hour by 60
\hourminute=\hour \multiply \hourminute by 60
\minute=\time \advance \minute by -\hourminute
\newcommand{\mydate}{\ \today \ - \number\hour :\number\minute}

%
%
\preprint{UMD-40762-407}

\title{\bf Isotropic and Anisotropic Lattice Spacing Corrections for I=2 $\pi\pi$ Scattering from Effective Field Theory}

\author{Michael I. Buchoff}
\email[]{mbuchoff@umd.edu}
\affiliation{Maryland Center for Fundamental Physics\\
	Department of Physics, University of Maryland,
	College Park, MD 20742-4111}

\date{\mydate}
%
%
\begin{abstract}
The calculation of the finite lattice spacing corrections for I=2 $\pi\pi$ scattering is carried out for isotropic and anisotropic Wilson lattice actions.  Pion masses and decay constants are also determined in this context.  These results correct the phase shift calculated from the lattice, which is connected to the scattering length and effective range in this low energy scattering process. When in terms of the lattice-physical parameters for either Wilson action, these lattice spacing effects first appear at the next-to-leading order counter-terms. 
\end{abstract}

\maketitle

%

%
%
\section{Introduction}

Numerical scattering calculations in lattice QCD are being performed by several collaborations.  These calculations are performed through the analysis of two hadrons in finite volume  \cite{Huang:1957im,Hamber:1983vu,Luscher:1990ux,Luscher:1986pf}.  One such scattering that has gained much attention in the field is I=2 $\pi\pi$ scattering.  Such numerical calculations (usually involving phase shifts and scattering lengths) have been calculated using Wilson lattice actions \cite{Gupta:1993rn,Fukugita:1994na,Fukugita:1994ve,Fiebig:1999hs,Aoki:1999pt,Liu:2001ss,Aoki:2001hc,Aoki:2002in,Aoki:2002sg,Aoki:2002ny,Ishizuka:2003nb,Yamazaki:2004qb,Du:2004ib,Aoki:2004wq,Aoki:2005uf,Li:2007ey} along with a several other lattice actions  \cite{Sharpe:1992pp,Kuramashi:1993ka, Juge:2003mr}.  Additionally, there are currently only two fully dynamical, 2+1 flavor calculations of I=2 $\pi\pi$ scattering, which use mixed lattice actions \cite{Beane:2005rj,Beane:2007xs} . Regardless of the action, unphysical lattice artifacts due to the finite lattice spacings exist in the numerical results of these calculations.   Therefore, measures should be taken in order to remove these effects so that the results from the lattice can best represent the continuum limit. The analysis in this paper is applicable for both isotropic and anisotropic Wilson actions.

Effective field theory (EFT)  provides a framework by which one can remove these unphysical effects.  Lattice spacing effects were first made explicit in chiral perturbation theory (\CPT) by Sharpe and Singleton \cite{Sharpe:1998xm}.  For the Wilson action, the chiral breaking terms that depend on the lattice spacing can be accounted for in a similar way to the chiral breaking quark mass.  Such methods have been extended to mixed-action, partially quenched theories for mesons through $\mc O(a)$ and $\mc O(a^2)$ \cite{Rupak:2002sm,Bar:2003mh,Chen:2006wf}, and baryons through $\mc O(a)$ \cite{Beane:2003xv} and $\mc O(a^2)$ \cite{Tiburzi:2005vy,Tiburzi:2005is,Chen:2007ug}. Additionally, Ref.~\cite{Chen:2006wf} carries out multiple meson scattering calculations (including I=2 $\pi\pi$ scattering) for mixed lattice actions and shows that for actions with chiral valence fermions, mesonic scattering parameters in terms of the lattice-physical parameters will have no counter-terms dependent on lattice spacing through next-to-leading order.   Alternatively, this work calculates these lattice spacing effects\footnote{These effects are in addition to the finite volume corrections to I=2 $\pi\pi$ scattering from Ref.~\cite{Bedaque:2006yi}} for I=2 $\pi\pi$ scattering for the chiral breaking Wilson fermions in both valance and sea sectors, and shows that these finite lattice spacing effects first appear in the next-to-leading order counter-terms for this action.

Many collaborations are now using anisotropic lattices (lattices with different temporal and spacial lattice spacings) as opposed the the usual isotropic lattices.  Such lattices can probe higher energy states (inverse time spacings $a_t^{-1} \sim 6$ GeV) and allow for a greater resolution (more data points).  However, anisotropic lattices lead to new lattice artifacts, including terms that explicitly break hypercubic symmetry.  Recent work has derived these anisotropic lattice artifacts in \CPT for $\mc O(a)$ and $\mc O(a^2)$ for mesons and baryons \cite{Bedaque:2007xg}.  There are several I=2 $\pi\pi$ scattering results published for anisotropic Wilson lattices \cite{Liu:2001ss,Du:2004ib,Li:2007ey}, which can benefit from removing these additional lattice artifacts.  

This paper presents the I=2 $\pi\pi$ scattering results from the isotropic \CPT and the anisotropic \CPT.  The pion mass and decay constant are also determined in this context. Sec.~\ref{sec:Lat_Scat} presents scattering on the lattice and defines the relevant quantity, $k \cot \d_0$, used to make comparisons between \CPT and the actual lattice calculation.  Next, in Sec.~\ref{sec:continuum}, the continuum scattering theory from \CPT is formulated in the context of this paper (originally worked out before \cite{Weinberg:1966kf,Gasser:1983yg,Bijnens:1997vq,Colangelo:2001df}).   Then, in Sec.~\ref{sec:iso}, the continuum result is extended for the isotropic Wilson lattice and finally, in Sec.~\ref{sec:aniso}, the result is extended for the anisotropic Wilson lattice.

%
%
\section{Scattering on the Lattice}\label{sec:Lat_Scat}

The Euclidean two-hadron correlation function in infinite volume gives no information about the Minkowski scattering amplitude (except at kinematic thresholds) \cite{Maiani:1990ca}.  However, when the correlation functions of two hadrons in a finite box are analyzed, the resulting energy levels are given by the sum of the energies of these two hadrons plus an additional energy of interaction, $\D E$, which is related to the scattering phase shift, $\d_l$ \cite{Huang:1957im,Hamber:1983vu,Luscher:1990ux,Luscher:1986pf}.   The $l$ subscript here represents the partial wave contribution of the phase shift.  In the infinite volume, the relation between the total scattering amplitude, $T(s,\th)$, and the partial waves amplitude, $t_l(s)$, is given by 

\begin{equation}\label{eq:Partial_Wave }
T(s,\th) = \sum_{l=0}^\infty (2l+1)P_l(\cos \th)t_l(s),
\end{equation}
where $s=4(\mp^2+k^2)$, and $k$ is the magnitude of the 3-momentum of the incoming particle in the center-of-mass frame.  The partial scattering amplitude $t_l(s)$ is related to the phase shift, $\d_l$ by

\begin{equation}\label{eq:t_to_delta }
t_l(s) = 32\pi \sqrt{\frac{s}{s-4\mp^2}}\frac{1}{2i}[e^{2i\d_l(s)}-1]=32\pi \sqrt{\frac{s}{s-4\mp^2}}\frac{1}{\cot \d_l - i}.
\end{equation}

These relations allow one to compare the calculated scattering amplitude (in \CPT) to the lattice calculation of $\d_l$.  The s-wave ($l = 0$) scattering amplitude is the dominant contribution to the total scattering amplitude in most low energy scattering processes and gives the cleanest signal in the lattice calculation.   The s-wave projection of the continuum scattering amplitude, $t_0(s)$, is

\begin{equation}\label{s_proj}
t_0(s) = \frac{1}{2}\int_{-1}^1 T(s,\th)  \: d(\cos \th).
\end{equation}

This s-wave scattering amplitude will be the scattering amplitude analyzed throughout the rest of this paper.  Following the discussion in Ref.~\cite{Bedaque:2006yi}, through one loop order in perturbation theory in Minkowski space, $t_0(s)$ can be written as

\begin{equation}\label{t0_bubbles}
t_0(s) \simeq t_0^{(LO)}(s) + t_0^{(NLO,R)}(s) + i t_0^{(NLO,I)}(s) \simeq \frac{(t_0^{(LO)}(s))^2}{t_0^{(LO)}(s) - t_0^{(NLO,R)}(s) - i t_0^{(NLO,I)}(s)},
\end{equation}
where $t_0^{(LO)}(s)$ is the leading order s-wave scattering amplutide, and $t_0^{(NLO,R)}(s)$ ($t_0^{(NLO,I)}(s)$) is the real (imaginary) part of the next-to-leading order s-wave scattering amplitude.  At this point, it is advantageous to introduce a $K$-matrix, which is defined through one loop as

\begin{equation}\label{k_mat}
K(s)  \equiv \frac{(t_0^{(LO)}(s))^2}{t_0^{(LO)}(s) - t_0^{(NLO,R)}(s)} .
\end{equation}

Taking the real part of the reciprocal of Eq.~\eqref{eq:t_to_delta } and Eq.~\eqref{t0_bubbles} and comparing to Eq.~\eqref{k_mat}, one gets the relation

\begin{equation}\label{k_rel}
\frac{1}{K(s)} = Re\bigg(\frac{1}{t_0(s)}\bigg) = \frac{1}{32\pi}\sqrt{\frac{s-4\mp^2}{s}}\cot\d_0(s), 
\end{equation}
where

\begin{equation}\label{re_rec_t}
Re\bigg(\frac{1}{t_0(s)}\bigg) = \frac{Re \;  t_0(s)}{\big(Re \;  t_0(s)\big)^2 + \big(Im \;  t_0(s)\big)^2} \approx \frac{1}{t_0^{(LO)}(s)}\bigg(1-\frac{t_0^{(NLO,R)}(s)}{t_0^{(LO)}(s)} \bigg).
\end{equation}
It is worth noting that when keeping terms though one loop, $Im \; t_0(s)$ does not contribute (it contributes at the next order).  Combining Eq.~\eqref{k_mat}, Eq.~\eqref{k_rel}, and Eq.~\eqref{re_rec_t}, one arrives at the continuum result

\begin{equation}\label{kcot_cont_amp}
k\cot\d_0(s) = 16\pi \sqrt{s} \: Re\bigg(\frac{1}{t_0(s)}\bigg) \approx 16\pi \sqrt{s}\frac{1}{t_0^{(LO)}(s)}\bigg(1-\frac{t_0^{(NLO,R)}(s)}{t_0^{(LO)}(s)} \bigg).
\end{equation}

As previously mentioned, lattice scattering calculations are performed in Euclidean space at finite volume.  The Euclidean amputated four-point correlator from the lattice, $\mc \tau_0(s)$,  is given by

\begin{equation}\label{lat_amp}
\mc \tau_0(s) \simeq \frac{(t_0^{(LO)}(s))^2}{t_0^{(LO)}(s) - t_0^{(NLO,R)}(s)-\D t_0(s) - \frac{(t_0^{(LO)})^2}{16\pi^2 L \sqrt{s}}\mc S \Big(\frac{(s-4\mp^2) L^2}{4 \pi^2} \Big)},
\end{equation}
where $\D t_0(s)$ represents all of the non-physical lattice artifacts (lattice spacing errors, finite volume errors ,etc.), $s$ is related to the energy of interaction, $\D E$, and $\mc S$ is a universal function of $s$ \cite{Luscher:1990ux,Beane:2003yx,Beane:2003da}.  If both pions in the box start with no external momentum, then $s=(\D E +2m_\pi)^2$.  In this paper, the only effect from lattice artifacts that will be included in $\D t_0$ is the lattice spacing effect.  Manipulating Eq.~\eqref{lat_amp}:

\begin{align}\label{lat_amp_pole}
\mc \tau_0(s) &\simeq \frac{1}{\frac{1}{K(s)}-\frac{\D t_0(s)}{(t_0^{(LO)})^2} - \frac{1}{16\pi^2 L \sqrt{s}}\mc S \Big(\frac{(s-4\mp^2) L^2}{16 \pi^2} \Big)} \nonumber\\
&=\frac{16\pi\sqrt{s}}{k\cot \d_0(s) - 16\pi \sqrt{s} \frac{\D t_0(s)}{(t_0^{(LO)})^2} - \frac{1}{\pi L}\mc S \Big(\frac{(s-4\mp^2) L^2}{16 \pi^2} \Big)}. 
\end{align}

The energy states are given by the poles of Eq.~\eqref{lat_amp_pole}, which are given by \cite{Bedaque:2006yi}

\begin{equation}\label{eq:LR }
k\cot \d_0 + \D(k\cot \d_0) = \frac{1}{\pi L} \mc S \bigg(\frac{(s-4\mp^2)L}{16\pi^2}\bigg),
\end{equation}
where
\begin{equation}\label{eq:D_t }
\D(k\cot \d_0) = -16\pi \sqrt{s}  \frac{\D t_0(s)}{(t_0^{(LO)})^2}.
\end{equation}

In general, most lattice calculations give their results in terms of the scattering length, $a_{\pi\pi}^{I=2}$.  One can extract the scattering length and the effective range, $r_{\pi\pi}^{I=2}$,  via the expansion of $k\cot \d_0$:
\begin{equation}\label{eq:cot_expand }
k\cot \d_0 = \frac{1}{a_{\pi\pi}^{I=2}} + \frac{1}{2}r_{\pi\pi}^{I=2}k^2 + \cdots.
\end{equation}

It is important to note that the prescription given above for finding the scattering length and effective range implies that the lattice artifacts, $\D(k\cot \d_0)$, have already been subtracted \textit{before} the expansion.  In this paper, continuum results are given in terms of $k\cot \d_0$ and lattice artifacts are given in terms of $\D(k\cot \d_0)$.

For results given in terms of the scattering length and effective range, one can relate Eq.~\eqref{eq:cot_expand } to the left hand side of Eq.~\eqref{eq:LR } to arrive at

\begin{equation}\label{eq:cot_expand_art }
k\cot \d_0+\D(k\cot \d_0) = \bigg(\frac{1}{a_{\pi\pi}^{I=2}}+\D\Big(\frac{1}{a_{\pi\pi}^{I=2}}\Big)\bigg) + \frac{1}{2}\bigg(r_{\pi\pi}^{I=2}+\D r_{\pi\pi}^{I=2} \bigg)k^2 + \cdots,
\end{equation}
where 

\begin{equation}\label{eq:recip_a_art }
\D\Big(\frac{1}{a_{\pi\pi}^{I=2}}\Big)= \D(k\cot \d_0)|_{k^2=0},
\end{equation}
and

\begin{equation}\label{eq:r_art }
\D r_{\pi\pi}^{I=2} = 2 \frac{d\big(\D(k\cot \d_0)\big)}{dk^2}\bigg|_{k^2=0}.
\end{equation}

While these relations are not too complicated, they do add additional steps to the calculation when compared to working with only $k\cot \d_0$ and $\D(k\cot \d_0)$.  Therefore, if one wants to extract the scattering length and effective range from the lattice calculation, one should first subtract  $\D(k\cot \d_0)$ from the right hand side of Eq.~\eqref{eq:LR } and then expand to determine the individual parameters\footnote{Current numerical calculations can only determine $k\cot \d_0$ for a limited number of $k$ values.  This leads to inaccuracies in expansions of $k^2$ and adds difficulty to finding the effective range.}.  This paper relates $k\cot \d_0$ and $\D(k\cot \d_0)$ to the effective field theory of the lattice.

%
%
\section{Chiral Perturbation Theory Results for $k\cot \d_0$ and $\D(k\cot \d_0)$}\label{sec:t_dt_CPT}

In leading order (LO) and next-to-leading-order (NLO) chiral perturbation theory (\CPT), it is necessary to introduce several undetermined low energy constants (LECs) in order properly account for corrections and counter-terms.   The number of independent LECs in the continuum depends on whether there are two flavors or more.  For I=2 $\pi\pi$ scattering being calculated here, only two flavor \CPT ($SU(2)_L\otimes SU(2)_R$ chiral symmetry) is needed for extrapolation.

%
%
\subsection{Continuum}\label{sec:continuum}

From the continuum \CPT Lagrangian, one can predict numerous results for different low energy processes involving hadrons.  However, the extent of the accuracy of these predictions are ultimately tied to how well the LECs are known.  For this reason, there has been much effort in the lattice community to try to determine these values \cite{Bernard:2006zp,Boucaud:2007uk,DelDebbio:2006cn,DelDebbio:2007pz,Leutwyler:2007ae}.

The continuum Lagrangian in \CPT is determined order by order in $Bm_q$ and $p^2$.  The Lagrangian through $\mc O(p^4)$ for two flavors is given by \cite{Gasser:1983yg}

\begin{align} \label{eq:L_cont}
    \mc L_{cont} =& \frac{f^2}{8}\tr(\delmd \S \delmu \S^\dag) + \frac{Bf^2}{4}\tr(m_q \S^\dag + \S m_q)  \nonumber\\
    &+\frac{\ell_1}{4}\big[\tr(\delmd \S \delmu \S^\dag)\big]^2+ \frac{\ell_2}{4} \tr(\delmd \S \delnd \S^\dag) \tr(\delmu \S^\dag \delnu \S) \nonumber\\
    &+\frac{(\ell_3+\ell_4)B^2}{4}\big[\tr(m_q \S^\dag + \S m_q)\big]^2 + \frac{\ell_4B}{4}\tr(\delmd \S \delmu \S^\dag)\tr(m_q \S^\dag + \S m_q) ,
\end{align}
where $f \sim 132$ MeV, $\ell_{1-4}$ are the original Gasser-Leutwyler coefficients defined in Ref.~\cite{Gasser:1983yg} and

\begin{align}\label{eq:def}
    &\S = \exp \Big(\frac{2i\phi}{f}\Big)\, ,&
    &\phi = \left(\begin{array}{cc}\frac{\pi_0}{\sqrt{2}} & \pi^+ \\ \pi^- & -\frac{\pi_0}{\sqrt{2}}\end{array}\right)\, ,&
    &m_q = \left(\begin{array}{cc}\bar{m} & 0 \\0 & \bar{m}\end{array}\right).
\end{align}
At LO, the resulting condensate is

\begin{equation}\label{qq_cond }
B = \lim_{m_q\rightarrow 0} \frac{ | \langle \bar{q} q \rangle |}{f^2}. 
\end{equation}

From  Eq.~\eqref{eq:L_cont}, one can calculate the physical values for the mass of the pion  ($\mp$) and the pion decay constant ($\fp$) to LO and NLO.  These expressions are given by

\begin{eqnarray}\label{eq:cont_f_m}
    \mp^2 &=& m^2 + \frac{1}{3f^2}[4\mp^2i\mc I(\mp) - m^2i\mc I(\mp)] + 4\ell_3 \frac{m^4}{\fp^2} \\
    &&\nonumber \\
    \fp &=& f \Big [1-\frac{2}{\fp^2}i\mc I(\mp) + 2\ell_4\frac{m^2}{\fp^2} \Big ] 
\end{eqnarray}
where $m^2$ and $\mc I(\mp)$ are defined below in Eq.~\eqref{eq:int_def}.  When evaluating the scattering amplitude from \CPT, one has the option of either expressing the answer in terms of the bare parameters ($f$ and $m$) or in terms of lattice-physical parameters ($\fp$ and $\mp$).  Throughout this paper, the bare parameters will always be eliminated from the scatting amplitude.  The continuum I=2 $\pi\pi$ scatting length at infinite volume is given by

\begin{align}\label{eq:cont_scatt_amp}
    T_{cont}= -\frac{2}{\fp^2} \bigg\{ s &- 2\mp^2 - \frac{2(3s-4\mp^2)}{3\fp^2}i\mc I(\mp) + \frac{(s-2\mp^2)^2}{\fp^2}i\mc J(\mp,p_s) \nonumber \\
    &+\frac{1}{3\fp^2}\big [ 3(t^2-\mp^4)+t(t-s)-2t\mp^2+4s\mp^2-2\mp^4 \big ] i\mc J(\mp,p_t) \nonumber \\
    &+ \frac{1}{3\fp^2}\big [ 3(u^2-\mp^4)+u(u-s)-2u\mp^2+4s\mp^2-2\mp^4 \big ] i\mc J(\mp,p_u) \nonumber \\
    &-\frac{1}{9(4\pi\fp)^2}\big [ 2s^2+6s\mp^2 -8\mp^4 -t^2 - u^2 \big] \nonumber \\
    &-\frac{4\ell_1}{\fp^2} \big[(t -2\mp^2)^2 + (u -2\mp^2)^2 \big ] \nonumber\\
    &- \frac{2\ell_2}{\fp^2}\big [2(s-2\mp^2)^2 + (t-2\mp^2)^2 + (u-2\mp^2)^2 \big] \nonumber \\
    &-8\ell_3 \frac{\mp^4}{\fp^2} + 4\ell_4 \frac{\mp^2(s-2\mp^2)}{\fp^2} \bigg\}
\end{align}
where
\begin{eqnarray}\label{eq:int_def}
m^2 & = & 2B\bar{m} \nonumber\\
\mc I(\mp) & = & \int_\mathcal{R} \frac{d^4k}{(2\pi)^4}\frac{1}{k^2-m^2} \nonumber\\
\mc J(\mp,P) & = & \int_\mathcal{R} \frac{d^4k}{(2\pi)^4}\frac{1}{[(k+P)^2-m^2]}\frac{1}{[k^2-m^2]}. 
\end{eqnarray}

This scatting amplitude includes all the partial wave contributions (this $T$ is the same as the $T(s,\th)$ that appears in Eq.~\eqref{eq:Partial_Wave }).  When projecting on the s-wave, expanding through $\mc O(k^2/\mp^2)$ and using Eq.~\eqref{kcot_cont_amp}, the result for $k\cot \d_0$(for the regularization and renormalization scheme defined in Ref.~\cite{Bijnens:1997vq})  is

\begin{align}\label{cont_k}
k\cot \d_0 \approx -\frac{8\pi\fp^2}{\mp} \Bigg\{& \bigg(1-\frac{\mp^2}{(4\pi\fp)^2} \Big[3\ln\Big(\frac{\mp^2}{\mu^2} \Big) -1 + \ell_{\pi\pi}^a(\mu)\Big]\bigg) \nonumber\\
&-\frac{1}{2} \bigg(3+\frac{\mp^2}{(4\pi\fp)^2} \Big[\frac{17}{3}\ln\Big(\frac{\mp^2}{\mu^2} \Big) +\frac{31}{3} + \ell_{\pi\pi}^r(\mu)\Big]\bigg)\frac{k^2}{\mp^2}+ \cdots \Bigg\},
\end{align}

where $\ell_{\pi\pi}^a(\mu)$ and  $\ell_{\pi\pi}^r(\mu)$ are linear combinations of the Gasser-Leutwyler coefficients given by \cite{Bijnens:1997vq}
\begin{align}\label{eq:def_l}
    \ell_{\pi\pi}^a(\mu) =& -4(4\pi)^2\Big(4\big(\ell_1^R(\mu) + \ell_2^R(\mu) \big) +\big(\ell_3^R(\mu)-\ell_4^R(\mu)\big)\Big),\nonumber\\
    \ell_{\pi\pi}^r(\mu) =& 4(4\pi)^2\big(12\ell_1^R(\mu)+4\ell_2^R(\mu)+7\ell_3^R(\mu)-3\ell_4^R(\mu)\big).
\end{align}
The superscript $R$ represents the renormalized Gasser-Leutwyler coefficients with scale-dependence.

In order for these continuum predictions from \CPT to be useful in a physical context, one needs to determine the values for $\ell_{\pi\pi}^a(\mu)$ and  $\ell_{\pi\pi}^r(\mu)$, which are undetermined from \CPT alone. While numerous values have been quoted for the Gasser-Leutwyler coefficients \cite{Leutwyler:2007ae}, it is still beneficial to determine these values with more precesion.  Therefore, it is prudent to use various lattice calculations at different pion masses determine these values.  However, since the lattice calculations are performed with discretized space and time, one needs to remove these lattice artifacts to extract the continuum result.

%
%
\subsection{Isotropic Discretization}\label{sec:iso}

To calculate finite lattice spacing corrections to I=2 $\pi\pi$ scattering for the isotropic Wilson action, one can follow the same steps done in the continuum case, but starting from a Lagrangian which includes these lattice spacing artifacts. The analysis on lattice spacing effects was done for the Symanzik action by Sheikholeslami and Wohlert \cite{Sheikholeslami:1985ij}.  From this analysis, the Lagrangian was made explicit in \CPT by Sharpe and Singleton \cite{Sharpe:1998xm}, followed by B\"ar, Rupak, and Shoresh \cite{Rupak:2002sm,Bar:2003mh}.  The power counting they used for this \CPT Lagrangian is
\begin{equation}\label{eq:pow_iso }
a_sW \sim Bm_q \sim p^2 \sim \e,
\end{equation}
where $a_s$ is the lattice spacing (same spacing in space and time direction) and $W$ is a condensate defined below.  The simplified two flavor Lagrangian to $\mc O(\e^2)$ is
     
\begin{align} \label{eq:L_iso}
    \mc L_{iso} =& \mc L_{cont}+ \D\mc L_{iso}.
\end{align}
\begin{align} \label{eq:dl_iso}
    \D\mc L_{iso} =& \frac{a_sWf^2}{4}\tr(\S^\dag + \S) +\frac{(w_3+w_4)a_sWB_0}{4}\tr(m_q \S^\dag + \S m_q)\tr(\S^\dag + \S) \nonumber\\  
    &+\frac{w_3^\prime (a_sW)^2}{4}\big[\tr(\S^\dag + \S)\big]^2+ \frac{w_4a_sW}{4}\tr(\delmd \S \delmu \S^\dag)\tr(\S^\dag + \S)
\end{align}

This Lagrangian is similar to Eq.~\eqref{eq:L_cont} with one new term at LO and three new terms at NLO.  All new terms are proportional to $a_s$ or $a_s^2$, which will vanish in the continuum limit as $a_s \rightarrow 0$.  At LO, there is a new condensate given by
\begin{equation}\label{SW_cond }
W = \lim_{m_q \rightarrow 0}c_{SW} \frac{ \langle \bar{q} \sigma_{\mu\nu} F^{\mu\nu} q \rangle}{f^2}.
\end{equation}
The new LECs at NLO are given by $w_3$, $w_3^\prime$, and $w_4$.  All these new terms obey the same symmetries as the Lagrangian in the continuum case and break chiral symmetry in a similar way to the quark mass.  It should also be noted that these new LECs depend on $a_s\ln a_s$ as well (as opposed to the Gasser-Leutwyler coefficients that have no dependence on the mass term). 

Furthermore, Ref.~\cite{Aoki:2007es} showed that the axial current (needed to calculate $\fp$) has an additional term at this order in \CPT given by 
\begin{eqnarray}
    \D A^a_\mu = 2aw_A \partial_\mu \tr\big(\s^a(\S-\S^\dag)\big).
\end{eqnarray}
This term (which can also be inferred from Ref.~\cite{Sharpe:2004ny}) leads to modifications of the LECs as well as the coefficient in front of the chiral logarithm in $\fp$.  Thus, this coefficient has dependence on the lattice artifacts at NNLO.  Ref.~\cite{Aoki:2007es} also points out that the condition for fixing the renormalization factor, $Z_A$, of the lattice currents needs to be mapped onto \CPT\footnote{The condition for fixing $Z_A$ is chosen by individual lattice calculations.  Ref.~\cite{Aoki:2007es} shows an example of this in \CPT by using the chiral Ward identities in infinite volume, which leads to  $\fp$ being free  lattice artifacts until NNLO.}.
   From this Lagrangian, $\fp$ and $\mp$ through NLO are \cite{Rupak:2002sm,Bar:2003mh,Aoki:2007es}

\begin{eqnarray}
    \mp^2 &=& (m^2 + 2a_sW) + \frac{1}{3f^2}[4\mp^2i\mc I(\mp) - (m^2+2a_sW)i\mc I(\mp)]\nonumber \\
    &&+ 4\ell_3 \frac{m^4}{\fp^2}+ 8w_3 \frac{a_sWm^2}{\fp^2}+16w_3^\prime \frac{(a_sW)^2}{\fp^2} \label{eq:iso_m}\\
    &&\nonumber \\
    \fp &=& f \Big [1-\frac{2}{\fp^2}i\mc I(\mp) + 2\ell_4\frac{m^2}{\fp^2}+4w_{eff}\frac{a_sW}{\fp^2} \Big ], \label{eq:iso_f}
\end{eqnarray}
where $w_{eff}$ in $\fp$ includes $w_4$ and $w_A$ and can vary based on the given renormalization condition for the axial current.  

To acquire the continuum result of these quantities, one needs to remove all the terms with dependence on $a_s$ or $a_s^2$.   The resulting I=2 $\pi\pi$ scatting amplitude with the physical parameters restored is given by

\begin{equation}\label{eq:T_iso}
T_{iso}=T_{cont}+\D T_{iso},
\end{equation}
where the $\mp$ and $\fp$ in $T_{cont}$ are given by Eq.~\eqref{eq:iso_m} and Eq.~\eqref{eq:iso_f} and $\D T_{iso}$ is given by

\begin{align}\label{eq:iso_scatt_amp}
    \D T_{iso}= -\frac{2}{\fp^2} \bigg\{&-16(w_3-2\ell_3)\frac{a_sW\mp^2}{\fp^2} -32(w_3^\prime-w_3+\ell_3)\frac{(a_sW)^2}{\fp^2}\nonumber\\
    &+8(w_{eff}-\ell_4)\frac{a_sW(s-2\mp^2)}{\fp^2}\bigg\}.
\end{align}

It is worth noting that by restoring the physical parameters, $m^2$ does not appear, and $a_sW$ only appears in the terms containing the LECs (the $\ell$ and $w$ terms).  This is a bit different than the continuum case where one could eliminate $m^2$ with only $\mp^2$.  Now, one eliminates $m^2$ with $(\mp^2-2a_sW)$, and thus, several LECs are multiplied by factors of $a_sW$.  In addition, all of the continuum results without LECs remain unchanged since each vertex will only contribute $\mp^2$ when the physical parameters are restored.  Using the relation in Eq.~\eqref{eq:D_t }, the resulting artifact for the isotropic Wilson lattice, $\D(k\cot \d_0)_{iso}$ is given by

\begin{align}\label{iso_k}
\D(k\cot \d_0)_{iso} \approx \frac{\mp}{2\pi} \Bigg\{& \bigg(w_{\pi\pi}^a(\mu)\frac{a_sW}{\mp^2}+w_{\pi\pi}^{\prime a}(\mu)\frac{(a_sW)^2}{\mp^4}\bigg) \nonumber\\
&-\frac{1}{2} \bigg(w_{\pi\pi}^r(\mu)\frac{a_sW}{\mp^2}+7w_{\pi\pi}^{\prime a}(\mu)\frac{(a_sW)^2}{\mp^4}\bigg)\frac{k^2}{\mp^2}+ \cdots \Bigg\},
\end{align}
where

\begin{align}\label{eq:def_w}
    w_{\pi\pi}^a(\mu) =& -8(4\pi)^2\big(w_3^R(\mu)-w_{eff}^R(\mu) -2\ell_3^R(\mu)+\ell_4^R(\mu)\big),\nonumber\\
    w_{\pi\pi}^{\prime a}(\mu) =&-16(4\pi)^2\big(w_3^{\prime R}(\mu)-w_3^R(\mu)+\ell_3^R(\mu)\big),\nonumber\\
    w_{\pi\pi}^r(\mu) =&-8(4\pi)^2\big(7w_3^R(\mu)-3w_{eff}^R(\mu)-14\ell_3^R(\mu)+3\ell_4^R(\mu)\big).
\end{align}

As seen in the results, the artifacts from the isotropic lattice that are present in the final form are either linear or quadratic in $a_s$.  By using results that differ in lattice spacing, one can pick off the coefficients of these artifacts and remove them from the final result.  If one is working with a perfectly clover-improved Wilson lattice, this would remove all $\mc O(a_s)$ effects leaving only the $\mc O(a_s^2)$ effects.  It is also important to note that there is no physical information gained by determining specific LECs that are a result of the isotropic lattice spacing (the individual $w$ terms) unlike determining specific Gasser-Leutwyler coefficents.  Therefore, the useful coefficient to extract is the linear combination of these terms so they can be removed from the final result. 

At this point, one can compare these lattice spacing effects for the Wilson action to those found for the mixed action (with chiral valence fermions) in Ref.~\cite{Chen:2005ab,Chen:2006wf}.  For this mixed action case, when in terms of the lattice-physical parameters, there is no lattice spacing dependence through the NLO counter-terms.  In contrast, when the lattice-physical parameters are restored in the Wilson action, these effects first appear at the NLO counter-terms.  Thus, these additional effects that are not present at NLO in the mixed action calculation will need to be removed for the Wilson action calculation.

%
%
\subsection{Anisotropic Discretization}\label{sec:aniso}

With the recent formulation of \CPT for the anisotropic lattice \cite{Bedaque:2007xg}, one can begin calculating corrections to various quantities of interest on the lattice.  This process is, in general, very similar to calculations in the continuum and isotropic lattices, but one now picks up additional terms that are a result of having different spacial and temporal spacings.  To help extract these effects in a more simplistic notation, the superscript $\xi$ has been added to all the new terms resulting from this anisotropy.  In practice, the anisotropic lattice picks up two new non-perturbative parameters: the parameter $\xi = a_s/a_t$ which is the measure of anisotropy and the parameter $\nu$, which is used to correct the ``speed of light" \cite{Klassen:1998fh,Chen:2000ej,Umeda:2003pj}.  By setting both parameters to 1, the isotropic limit is recovered.  In addition to the $W$ condensate defined in Eq.~\eqref{SW_cond }, we pick up a $W^\xi$ condensate that is given by
\begin{equation}\label{SW_xi_cond }
W^\xi = \lim_{m_q \rightarrow 0}c_{SW}^\xi (u^\xi)^\mu (u^\xi)_\nu \frac{ \langle \bar{q} \sigma_{\mu\l} F^{\nu\l} q \rangle}{f^2}.
\end{equation}
where $u_\mu^\xi$ is a vector that breaks hypercubic invariance.  It is important to note that the convention chosen for $u_\mu^\xi$ appears in the anisotropic \CPT and its observables.  For convenience, we choose this vector to be  $u_\mu^\xi = (1,\textbf{0})$. The condensates and the anisotropic paramaters are related at the classic level by (with Wilson coefficients $r_s=r_t=1$) 
\begin{eqnarray}\label{W_prop}
    W&\propto& c_{SW} \propto \nu,   \\
    W^\xi &\propto& c_{SW}^\xi \propto \frac{1}{2}\bigg(\frac{a_t}{a_s}-\nu \bigg).  
\end{eqnarray}
In the isotropic limit when $\xi$ and $\nu$ are set to 1, the isotropic condensate will remain and the anisotropic condensate will vanish.  Using a similar notation throughout, all terms that appear with a $\xi$ superscript will vanish when $a_s = a_t$ and $\nu = 1$.

The power counting convention used in Eq.~\eqref{eq:pow_iso } for the anisotropic Lagrangian is

\begin{equation}\label{eq:pow_aniso }
a_sW \sim a_sW^\xi \sim Bm_q \sim p^2 \sim \e.
\end{equation}
Writing this new Lagrangian in the form of Eq.~\eqref{eq:L_iso}, the two-flavor anisotropic \CPT Lagrangian through $\mc O(\e^2)$ is
\begin{align} \label{eq:L_aniso}
    \mc L_{aniso} =& \mc L_{cont}+ \D\mc L_{iso}+\D\mc L_{aniso}.
\end{align}
where

\begin{align} \label{eq:L_aniso}
    \D\mc L_{aniso} =& \frac{a_sW^\xi f^2}{4}\tr(\S^\dag + \S)+\frac{(w_3^\xi+w_4^\xi+w_1^\xi) a_sW^\xi B_0}{4}\tr(m_q \S^\dag + \S m_q)\tr(\S^\dag + \S)\nonumber\\ 
    &+\frac{\hat{w}_3^\xi (a_sW^\xi)^2}{4}\big[\tr(\S^\dag + \S)\big]^2+\frac{\bar{w}_3^\xi (a_sW)(a_sW^\xi)}{4}\big[\tr(\S^\dag + \S)\big]^2\nonumber\\
    &+ \frac{w_4^\xi a_sW^\xi}{4}\tr(\delmd \S \delmu \S^\dag)\tr(\S^\dag + \S)+\frac{w_1^\xi a_sW^\xi}{4}u^\mu u_\nu (\delmd \S \delnu \S^\dag)\tr(\S^\dag + \S).
\end{align}

In addition to the anisotropic condensate $W^\xi$ mentioned above at LO, there are five new LECs at NLO as a result of this anisotropy.  Four of these new LECs obey the same symmetry structure as the isotropic terms, however the $\w_1^\xi$ term additionally breaks hypercubic invariance.  Therefore, this term only corrects the time derivative, but not the spacial one (for the convention of $u_\mu^\xi$ chosen).  As a result, $\fp$ is parameterized by two constants; $\fpt$, which is $\fp$ measured in time, and $\fps$, which is $\fp$ measured in space.  This leads to one correction for the space-measured $\fps$ and a separate    correction for the $\fpt$.  The pion mass ($\mp$), the time-measured pion decay constant ($\fpt$), and the space-measured pion decay constant ($\fps$) through NLO are

\begin{eqnarray}
    \mp^2 &=& (m^2 + 2a_sW+2a_sW^\xi) + \frac{1}{3f^2}[4\mp^2i\mc I(\mp) - (m^2+2a_sW+2a_sW^\xi)i\mc I(\mp)]\nonumber \\
    &&+ 4\ell_3 \frac{m^4}{\fpt^2}+ 8w_3 \frac{a_sWm^2}{\fpt^2}+16w_3^\prime \frac{(a_sW)^2}{\fpt^2}\nonumber \\
    &&+8w_3^\xi \frac{a_sW^\xi m^2}{\fpt^2}+16\hat{w}_3^\xi \frac{(a_sW^\xi)^2}{\fpt^2} + 16\bar{w}_3^\xi \frac{(a_sW)(a_sW^\xi)}{\fpt^2}  \label{eq:aniso_m}\\
    &&\nonumber \\
    \fpt &=& f \Big [1-\frac{2}{\fpt^2}i\mc I(\mp) + 2\ell_4\frac{m^2}{\fpt^2}+4w_4\frac{a_sW}{\fpt^2} +4(w_{eff}^\xi+w_1^\xi)\frac{a_sW^\xi}{\fpt^2} \Big ]  \label{eq:aniso_f}\\
    &&\nonumber \\
    \fps &=& f \Big [1-\frac{2}{\fpt^2}i\mc I(\mp) + 2\ell_4\frac{m^2}{\fpt^2}+4w_4\frac{a_sW}{\fpt^2} +4w_{eff}^\xi\frac{a_sW^\xi}{\fpt^2} \Big ],
\end{eqnarray}
where, as in the isotropic case, the $w_{eff}^\xi$ depends on the  renormalization condition for the axial current (this LEC is the same for both $\fpt$ and $\fps$).

For the rest of this section, all calculations are given in terms of $\mp$ and $\fpt$.  As mentioned before, how one accounts for the effect of this hypercubic breaking term depends on the convention.  In the convention used here, only $\fpt$ sees the effect of this term and $\fps$ does not.  The scatting amplitude is given by

\begin{equation}\label{ }
T_{iso}=T_{cont}+\D T_{iso}+\D T_{aniso}
\end{equation}
where the $\mp$ and $\fpt$ in $T_{cont}$ are given by Eq.~\eqref{eq:aniso_m} and Eq.~\eqref{eq:aniso_f} and $\D T_{aniso}$ is given by

\begin{align}\label{eq:aniso_scatt_amp}
    \D T_{aniso}= -\frac{2}{(\fpt)^2} \bigg\{&-16(w_3^\xi-2\ell_3)\frac{a_sW^\xi\mp^2}{(\fpt)^2} -32(\hat{w}_3^\xi-w_3^\xi+\ell_3)\frac{(a_sW^\xi)^2}{(\fpt)^2}\nonumber\\
    &-32(\bar{w}_3^\xi-w_3-w_3^\xi+2\ell_3)\frac{(a_sW)(a_sW^\xi)}{(\fpt)^2}\nonumber\\
    &+8(w_{eff}^\xi+w_1^\xi-\ell_4)\frac{a_sW^\xi(s-2\mp^2)}{(\fpt)^2}-16w_1^\xi \frac{a_sW^\xi k^2}{(\fpt)^2}\bigg\}.
\end{align}

Most of the effects in this scattering amplitude are similar to the isotropic case, except now there are  also expansions in $a_sW^\xi$ in addition to the expansions in $a_sW$.  Thus, as expected, if all the anisotropic effects are removed, only the isotropic limit remains.  The only new symmetry breaking effect is the $w_1$ term which is not a hypercubic invariant term.  In other words, all of the hypercubic breaking due to anisotropy at this order is contained in this term.  However, it's effects in $\D T_{aniso}$ appear as just another contribution to the linear combination of the LECs in front of the term $a_sW^\xi$.  Therefore, it is difficult to determine the effect of the the hypercubic breaking term alone from $\D T_{aniso}$  since its effects will be mixed in with the other anisotropic LECs\footnote{The total hypercubic breaking effect would be more visible from the differences of $\fpt$ and $\fps$}.   The resulting artifacts for the anisotropic Wilson lattice are the isotropic artifacts, $\D(k\cot \d_0)_{iso}$, and the anisotropic artifacts, $\D(k\cot \d_0)_{aniso}$.  Therefore, the total effect of the lattice artifacts due to lattice spacings are 

\begin{equation}
\D(k\cot \d_0) = \D(k\cot \d_0)_{iso} + \D(k\cot \d_0)_{aniso}.
\end{equation}
The anisotropic lattice artifacts are given by

\begin{align}\label{aniso_k}
\D(k\cot \d_0)_{aniso} \approx \frac{\mp}{2\pi} &\Bigg\{ \bigg(w_{\pi\pi}^{\xi a}(\mu)\frac{a_sW^\xi}{\mp^2}+\hat{w}_{\pi\pi}^{ \xi a}(\mu)\frac{(a_sW^\xi)^2}{\mp^4}+ \bar{w}_{\pi\pi}^{\xi a}(\mu) \frac{(a_sW)(a_sW^\xi)}{\mp^4}\bigg) \nonumber\\
&-\frac{1}{2} \bigg(w_{\pi\pi}^{\xi r}(\mu)\frac{a_sW^\xi}{\mp^2}+7w_{\pi\pi}^{\prime \xi a}(\mu)\frac{(a_sW^\xi)^2}{\mp^4}+ 7\bar{w}_{\pi\pi}^{\xi a}(\mu) \frac{(a_sW)(a_sW^\xi)}{\mp^4}\bigg)\frac{k^2}{\mp^2}+ \cdots \Bigg\},
\end{align}
where

\begin{align}\label{eq:def_w}
    w_{\pi\pi}^{\xi a}(\mu) =& -8(4\pi)^2\big(w_3^{\xi R}(\mu) - w_{eff}^{\xi R}(\mu)+w_1^{\xi R}(\mu) -2\ell_3^R(\mu)+\ell_4^R(\mu)\big) ,\nonumber\\
    \hat{w}_{\pi\pi}^{ \xi a}(\mu) =&-16(4\pi)^2\big(w_3^{\prime R}(\mu)-w_3^R(\mu)+\ell_3^R(\mu)\big),\nonumber\\
    \bar{w}_{\pi\pi}^{\xi a}(\mu) =&-16(4\pi)^2\big(\bar{w}_3^{\xi R}(\mu)-w_3^R(\mu)-w_3^{\xi R}(\mu)+2\ell_3^R(\mu)\big), \nonumber\\
    w_{\pi\pi}^{\xi r}(\mu)=& -8(4\pi)^2\big(7w_3^{\xi R}(\mu)+9w_{eff}^{\xi R}(\mu)+w_1^{\xi R}(\mu)-14\ell_3^R(\mu)-9\ell_4^R(\mu)\big).
\end{align}
  
As a result, the ultimate effects of the anisotropic lattice on $\D(k\cot \d_0)$ are more terms that will require variation of $a_s$ and $a_t$ independently to fit.  Three more terms require fitting to correct the constant term in the expansion and one more term requires fitting for the $\mc O(k^2/\mp^2)$ term in order to remove it's effects.  As mentioned with the isotropic correction, no physical information is gained by picking off the individual anisotropic LECs. Analogous to the isotropic case, the anisotropic Wilson action first has lattice spacing dependence at the NLO counter-terms. The aim here is to determine these linear combinations and remove their effects from the lattice measurements.

%
%
\section{Discussion \label{sec:concl}}

In this work, I=2 $\pi\pi$ scattering was calculated from the \CPT for isotropic and anisotropic lattice spacings.  Also, connections between these \CPT calculations and the $k \cot \d_0$ value measured from lattice calculations were illustrated.  When $\D(k \cot \d_0)$ is given in terms of the lattice-physical parameters, these lattice spacing effects first appear at the NLO LECs and can be removed from the result of the lattice calculation.  However,  $\D(k \cot \d_0)$ has numerous undetermined linear combinations of LECs, which would need to be determined (by fitting several different lattice spacings) in order to successfully remove it from the lattice result. Therefore, as more lattice calculations of $\pi\pi$ scattering are completed (for both isotropic and anisotropic lattice spacings), these combinations of LECs can be determined better, which will result in a more accurate result after these lattice artifacts are removed.   

%
%
\begin{acknowledgments}
I would like to thank Paulo Bedaque and Andr\'e Walker-Loud for useful discussions.  This research was supported in part by the U.S. Dept. of Energy under grant no. DE-FG02-93ER-40762.

\end{acknowledgments}

%
%

\end{document}